\documentclass[graybox]{svmult}

\usepackage{mathptmx}       
\usepackage{helvet}         
\usepackage{courier}        
\usepackage{type1cm}        
\usepackage{makeidx}         
\usepackage{graphicx}        
\usepackage{multicol}        
\usepackage[bottom]{footmisc}

\makeindex             

\begin{document}

\title*{Biologically-Inspired Electronics with Memory Circuit Elements}
\author{Massimiliano Di Ventra and Yuriy V. Pershin}
\institute{Massimiliano Di Ventra \at Department of Physics, University of California San Diego,
La Jolla, CA 92093 USA, \\ \email{diventra@physics.ucsd.edu}
\and Yuriy V. Pershin \at Department of Physics and Astronomy and
USC Nanocenter, University of South Carolina, Columbia, SC 29208 USA, \\ \email{pershin@physics.sc.edu}}
%
%
\maketitle

\abstract{Several unique properties of biological systems, such as adaptation to natural environment, or of animals to learn patterns when appropriately trained, are features that are extremely useful, if emulated by electronic circuits, in applications ranging from robotics to solution of complex optimization problems, traffic control, etc. In this chapter, we discuss several examples of biologically-inspired circuits that take advantage of memory circuit elements, namely, electronic elements whose resistive, capacitive or inductive characteristics depend on their past dynamics. We provide several illustrations of what can be accomplished with these elements including learning circuits and related adaptive filters, neuromorphic and cellular computing circuits, analog massively-parallel computation architectures, etc. We also give examples of experimental realizations of memory circuit elements and discuss opportunities and challenges in this new field.}

\section{Introduction}
\label{sec:1}

Reproducing some of the features that are commonly found in living organisms, including - as the ultimate, and most sought after goal - the workings of the human brain, is what "artificial intelligence" is all about~\cite{book_AI}. However, even without aiming for such an ambitious target, there are several tasks that living organisms perform seamlessly and, when reproduced in electronic circuits, are of great benefit and help us solve complicated problems. The main idea of {\it biologically-inspired electronics} is thus in borrowing approaches used by biological systems  interacting with their environment and in applying them in diverse technological areas requiring, for example,  adaptation to changing inputs, analog solution of optimization problems -- one of the well-known approaches to this problem is the "ant-search algorithm"~\cite{Colorni91a,book_ants} --, associative memory and unlearning, binary and fuzzy logic, etc.

If we look closer, it is evident that if we want to reproduce any of these tasks with electronic circuits, two main characteristics have to be satisfied by some of the circuit elements or their combinations. These elements need to {\it i}) store information -- they have to have memory of the past -- and {\it ii}) be dynamical -- their states have to vary in time in response to a given input, preferably in a non-linear way. The latter requirement will help building a wide range of electronic circuits of desired functionality.

Circuits based on active elements (such as transistors) can clearly perform both tasks, however, at a high cost of power consumption, low density, and complexity. It would be much more desirable if we could combine the above features in single, passive elements, preferably with dimensions at the
nanometer scale, and hence comparable to -- or even smaller than -- biological storing and processing units, such as synapses and neurons.

Such elements do exist, and go under the name of
{\it memristive}, {\it memcapacitive} and {\it meminductive systems}, or collectively simply named {\it memelements}~\cite{diventra09a}. These are resistors, capacitors and inductors, respectively, whose state at a given time depends on the history
of applied inputs (e.g., charge, voltage, current, or flux) and states through which the system has evolved. As we have shown, these memelements provide an unifying description of materials and systems with memory~\cite{diventra11a}, in the sense that all two-terminal electronic devices based on memory materials and systems, when subject to time-dependent perturbations, behave simply as -- or as a combination of -- memristive, memcapacitive and meminductive systems, namely, dynamical non-linear circuit elements with memory~\cite{pershin11a}.

In this Chapter, we will show how the analog memory features of memelements are ideal to reproduce a host of processes typical of living organisms. We will review mainly work by the authors in various contexts, ranging from learning (adaptive) circuits to associative memory, and inherent massive parallelism afforded by networks of memelements. The Chapter is then organized as follows. Section \ref{sec:2} briefly reviews both the definition of these memory circuit elements and their main properties (a full account can be found in the original publications~\cite{chua71a,chua76a,diventra09a} and in our recent review paper~\cite{pershin11a}). Several experimental realizations exemplifying memristive, memcapacitive and meminductive systems are discussed in Sec. \ref{sec:exp}. Section \ref{sec:3} is devoted to biologically-inspired circuits based on memory circuit elements including simple adaptive circuits (Sec. \ref{sec:31}), neuromorphic circuits (Sec. \ref{sec:32}) and massively-parallel analog processing circuits (Sec. \ref{sec:33}). Concluding remarks are given in Sec. \ref{sec:4}.

\section{Definitions and properties}
\label{sec:2}

\begin{figure}[tb]
\centering \includegraphics[width=8cm]{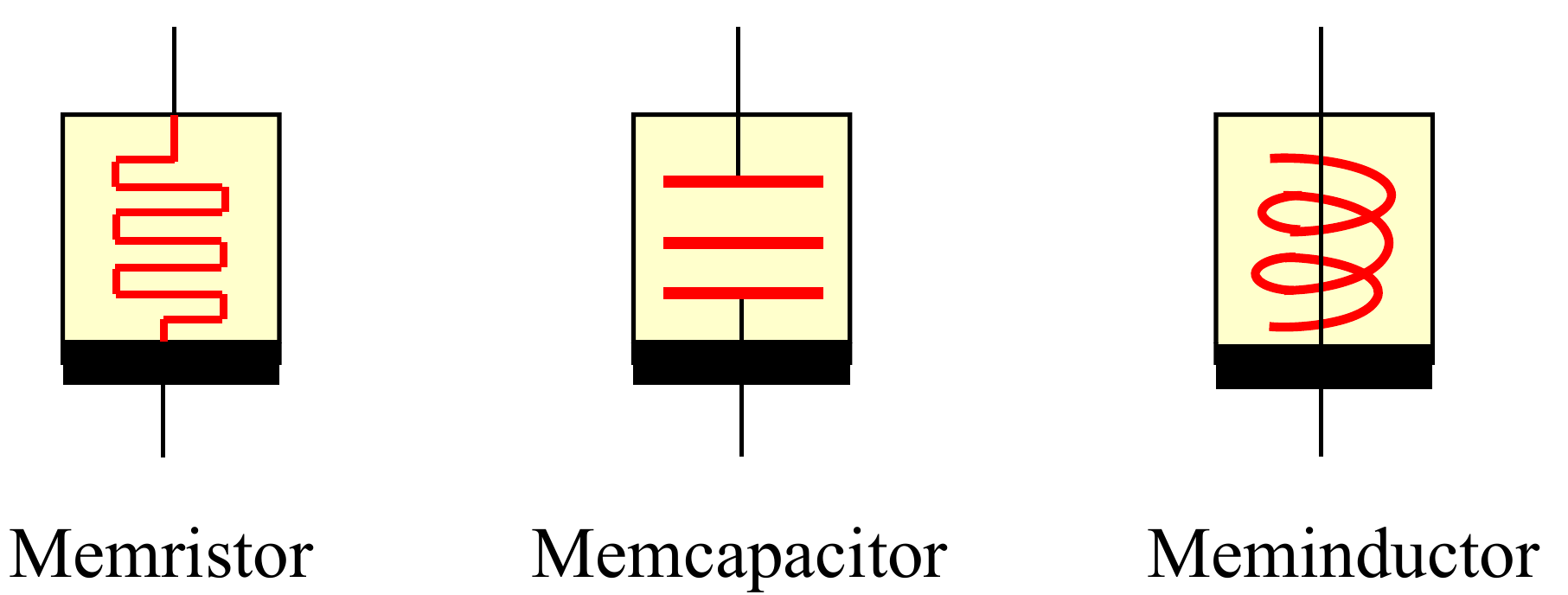}
\caption{Symbols of memory circuit elements: memristor, memcapacitor and meminductor. Generally, memelements are asymmetric devices. The thick horizontal lines above the bottom electrodes are employed to define the device polarity~\cite{diventra09a}. Reprinted with permission from Ref. \cite{diventra09a}. \copyright  2009 IEEE.}
\label{fig:1}       
\end{figure}

Let us consider electronic devices defined by all possible pairs of fundamental circuit variables $u(t)$ and $y(t)$ (i.e., current,
charge, voltage, or flux). For each pair, we can introduce a  response function, $g$, that, generally, depends on
a set of $n$ state variables, $x=\{x_i\}$, $i=1,..,n$, describing the internal state of the system~\cite{diventra09a} \footnote{There is no such
dependence for traditional basic circuit elements -- resistors, capacitors and inductors.}.
For instance, resistance of certain systems may depend on spin polarization~\cite{pershin08a,wang09a} or temperature~\cite{chua76a}; capacitance and inductance of some other elements may exhibit  dependence on system geometry \cite{pershin11c,pershin11a} or electric polarization~\cite{martinez09a}. In general, all internal microscopic physical properties related to the memory response of these electronic devices should be included into the vector of internal state variables $x$. The resulting memory circuit elements are then described by the following relations~\cite{diventra09a}
\begin{eqnarray}
y(t)&=&g\left(x,u,t \right)u(t) \label{Geq1}\\ \dot{x}&=&f\left(
x,u,t\right) \label{Geq2}
\end{eqnarray}
where $f$ is a continuous $n$-dimensional vector function. Eqs. (\ref{Geq1}), (\ref{Geq2})
have to be supplied by appropriate initial conditions~\cite{Corinto11a}. If $u$ is the current and $y$ is the voltage then Eqs. (\ref{Geq1}), (\ref{Geq2}) define memristive (for memory resistive) systems. In this case $g$ is the {\em memristance} (memory resistance). In memcapacitive (memory capacitive) systems, the
charge $q$ is related to the voltage $V$ so that $g$ is the {\em memcapacitance} (memory capacitance). Finally, in meminductive (memory inductive) systems the flux $\varphi$ is related to the current $I$ with $g$ the {\em meminductance} (memory inductance). There are still three additional pairs of fundamental circuit variables. However, these
do not give rise to any new devices. For example, the pairs charge-current and voltage-flux are linked
through equations of electrodynamics. Moreover, we could redefine devices defined by the charge-flux (which is the integral of the voltage) pair in the
current-voltage basis~\cite{chua71a}. Circuits symbols of memristive, memcapacitive and meminductive systems are presented in Fig. \ref{fig:1}.

Let us consider memristive systems in more details (definitions of all the other elements can be easily derived by considering the different constitutive variables \cite{diventra09a}). Specifically, a current-controlled
memristive system \cite{chua76a,diventra09a} is defined by Eqs. (\ref{Geq1}), (\ref{Geq2}) as
\begin{eqnarray}
V_M(t)&=&R\left(x,I,t \right)I(t), \label{eq1}\\
\dot{x}&=&f\left(x,I,t\right), \label{eq2}
\end{eqnarray}
where $V_M(t)$ and $I(t)$ denote the voltage and current across the
device, and $R$ is the memristance. In a simple model of an ideal memristor~\cite{chua71a}, the memristance depends only on charge -- the
time integral of the current. One particular realization of such a model has been suggested in Ref. \cite{strukov08a} and is formulated as
\begin{equation}
R_{ij}^M=R_{ON}x+R_{OFF}\left( 1 -x\right),\label{eq1struk}
\end{equation}
where $R_{ON}$ and $R_{OFF}$  are minimal and maximal values of memristance, and $x$ is a dimensionless internal state variable bound to the region $0\leq x \leq 1$. The dynamics of $x$ can then be simply
chosen as \cite{strukov08a}
\begin{equation}
\frac{\textnormal{d} x}{\textnormal{d} t}=\alpha I(t) \label{eq2struk},
\end{equation}
where $\alpha$ is a constant and $I(t)$  is the current flowing through the memristor.

Another example of memristive systems (which we will make use of later in this chapter) is a threshold-type memristive system~\cite{pershin09b}.
Its model is specified by a threshold voltage (and some other parameters) that defines different device response regions (with regard to the voltage applied across the device). Mathematically, the threshold-type memristive system is described by the following equations
\begin{eqnarray}
I&=&x^{-1}V_M, \label{eq3} \\ \dot x&=&\left(\beta V_M+0.5\left(
\alpha-\beta\right)\left[ |V_M+V_t|-|V_M-V_t| \right]\right)
\nonumber\\ & &\times \theta\left( x/R_1-1\right) \theta\left(
1-x/R_2\right) \label{eq4},
\end{eqnarray}
where $I$ and $V_M$ are the current through and the voltage drop on the device, respectively, and
$x$ is the internal state variable playing the role of memristance, $R=x$,
$\theta(\cdot)$ is the step function, $\alpha$  and $\beta$
characterize the rate of memristance change at $|V_M|\leq V_t$ and
$|V_M|> V_t$, respectively,
$V_t$ is a threshold voltage, and  $R_1$ and $R_2$ are limiting
values of the memristance $R$. In Eq. (\ref{eq4}), the $\theta$-functions
symbolically show that the memristance can change only between
$R_1$ and $R_2$. On a practical level, the value of $x$  must be monitored at each time step
and in the situations when $x<R_1$ or $x>R_2$, it must be set equal to
$R_1$ or $R_2$, respectively. In this way, we avoid situations
when $x$ may overshoot the limiting values by some amount and thus
not change any longer because of the step function in Eq.
(\ref{eq4}). We have introduced and employed this model to describe the learning properties of unicellular organisms, associative memory, etc., as we will describe in some detail in the following sections.

There are several properties that characterize memory circuit elements. We refer the reader to the original papers~\cite{chua71a,chua76a,diventra09a} and the extensive review~\cite{pershin11a} where these are discussed at length. Here,
we just mention that they are typically characterized by a frequency-dependent "pinched hysteresis
loop" in their constitutive variables when subject to a periodic input. Also, normally, the memristance, memcapacitance and meminductance acquire values between two limits (with exceptions as discussed in Refs.~\cite{pershin11a,martinez09a,krems2010a}). Although the hysteresis of these elements under a periodic input may strongly depend on initial conditions~\cite{Corinto11a}, it is generally more pronounced at frequencies of the external input that are comparable to frequencies of internal processes that lead to memory. In many cases, at very low frequencies memory circuit elements behave as non-linear elements while at high frequencies as linear elements.

Also, the hysteresis loops may or may not show self-crossing - which we have named type-I and type-II crossing behavior, respectively \cite{pershin11a} - and often the internal state variable remains unchanged for a long time without any input signal applied. This provides non-volatile memory, which is an important feature for some of the applications we discuss later.

Finally, we mention that the state variables - whether from a continuum or a discrete set of states - may follow a {\it stochastic differential equation} rather than a deterministic one \cite{pershin11a}. Interesting
effects have been predicted in the presence of noise, such as  noise-induced hysteresis \cite{Stotland11a}. This may have a large bearing in simulating biological processes - which necessarily occur under
 noisy conditions - and could be used to enhance the performance of certain devices.

\section{Experimental realizations}
\label{sec:exp}

\begin{figure}[tb]
\sidecaption
\includegraphics[width=7cm]{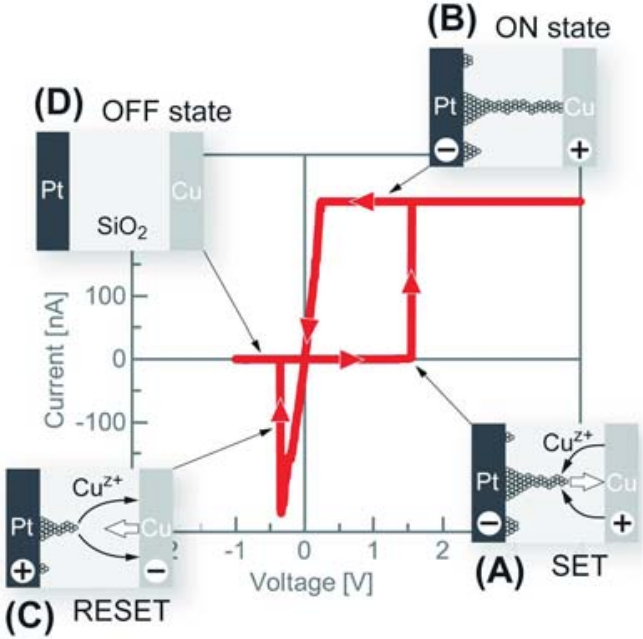}
\caption{Current-voltage characteristic of a Cu/SiO$_2$/Pt electrochemical metallization cell recorded using a triangular voltage sweep. The insets show dynamics of metallic filament formation.  (Reprinted with permission from \cite{Schindler09b}. © 2009 American Institute of Physics.) }
\label{fig:loop}       
\end{figure}

In this section, we briefly discuss experimental realizations of memory circuit elements. There is a large amount of experimental systems showing memristive behavior (based, however, on very different physical mechanisms). For example, in thermistors -- being among the first identified memristive systems~\cite{chua76a} -- the memory effects are related to thermal effects, mainly on how fast the heat dissipation occurs. In spintronics memristive systems, either based on semiconductor~\cite{pershin08a} or metal~\cite{wang09a} materials, the memory feature is provided by the electron spin degree of freedom. Finally, resistance switching memory cells are probably the most important type of memristive systems. These cells are normally built of two metal electrodes separated by a gap of a few tens of nanometers filled by a memristive material. Different physical processes inside the memristive material can be responsible for the memory. Fig. \ref{fig:loop} shows an example of resistance switching memory cell -- an electrochemical metallization cell -- in which a layer of dielectric material (SiO$_2$) separates two dissimilar metal electrodes (made of copper and platinum)~\cite{Schindler09b}. Externally applied voltage induces migration of copper atoms that can bridge the gap between the electrodes thus reducing the cell's resistance. Such a bridge can also be disrupted by the applied voltage of the opposite polarity (see Fig. \ref{fig:loop}).

Memcapacitive effects can be related to changes in the capacitor geometry or specific intrinsic properties of dielectric medium located between the capacitor plates~\cite{pershin11a}. For instance, the former mechanism plays the main role in elastic memcapacitive systems that could be based on internal elastic states of the capacitor plate (e.g., direction of bending~\cite{pershin11c}) or elasticity of a medium between the plates that could be thought of as a spring~\cite{pershin11a}. Examples of the latter mechanism include memcapacitive systems with a delayed polarization response~\cite{martinez09a,krems2010a} and structures with permittivity switching~\cite{Lai09a}. Although meminductive systems are  the least studied memory circuit elements at the moment, there are several known systems showing such type of functionality. For instance, in bimorph meminductive systems~\cite{lubecke01a,zine04a,chang06a}, the inductance depends on the inductor's shape defined by the inductor temperature. Heat dissipation mechanisms play a significant role in this type of systems. Many additional examples of memristive, memcapacitive and meminductive systems can be found in our recent review paper~\cite{pershin11a}. The most important aspect of all these examples is that they relate primarily to structures with nanoscale dimensions. This is not surprising since (up to a certain limit) the smaller the dimensions of the system the easier it is to observe memory effects.

In addition, we would like to mention that several types of memory effects may be present in a single device. For example, the coexistance of memristive and memcapacitive responses has been observed in perovskite oxide thin films~\cite{liu06a}. Moreover, several three-terminal transistor-like memristive devices have been investigated in the past~\cite{erokhin07a,Alibart10a,Lai10a}. Clearly, different types of memory materials can be combined to obtain multi-terminal device structures with complex functionalities.

\section{Biologically-Inspired Circuits}
\label{sec:3}

\subsection{Modeling adaptive behavior of unicellular organisms}\label{sec:31}

Adaptive behavior is common to all life forms on Earth of all five kingdoms of nature: in Plantae (the plants)~\cite{Kozlowski02a,Trewavas03a,Garzon11a}, Animalia (the animals)~\cite{book_adaptive}, Protista (the single-celled eukaryotes)~\cite{Ojal98a,saigusa08a}, Fungi (fungus and related organisms)~\cite{Li08a,Hartley97a}, and Monera (the prokaryotes)~\cite{Thieringer98a,Oost09a,Larsson11a}. (The literature on this subject is extensive, hence we
have given only a few representative references.)  To a greater or lesser extent, representatives of all life forms respond (adapt) to changes of their environment in a manner that increases the survival of their
species. It would thus be of benefit to mimic and use this important natural feature in artificial structures, in particular in electronics to allow novel functionalities otherwise nonexistent in standard circuitry. There are indeed
many domains where such {\it learning circuits} can be employed. These range from robot control systems to signal processing. Therefore, developing circuit models of the adaptive behavior of the natural world is of great
importance in many scientific areas~\cite{book_adaptive1}. Note that by "learning" here we simply mean the ability to adapt to incoming signals with retention of such information for later use.

\begin{figure}[bt]
\sidecaption
\includegraphics[width=5.5cm]{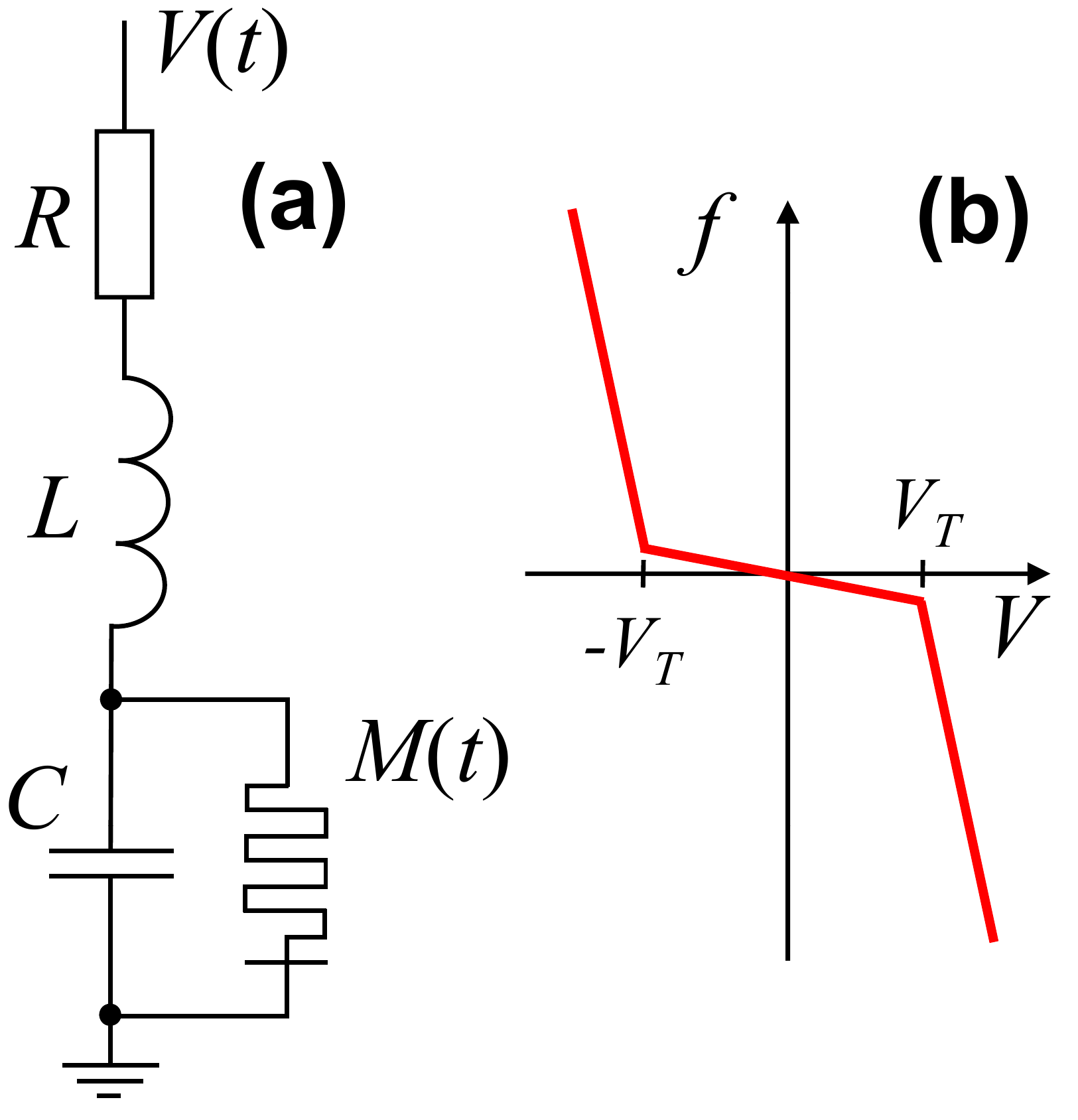}
\caption{(a) A possible realization of the learning circuit consisting of resistor $R$, inductor $L$, capacitor $C$ and memristive system $M$. (b) Schematics of the function $f(V)$ defining a threshold-type memristive system (see also Eqs. (\ref{eq3}), (\ref{eq4})). Reprinted with permission from Ref. \cite{pershin09b}. \copyright  2009 American Physical Society.}
\label{fig:2}       
\end{figure}

In this section we consider a particularly interesting example: the ability of the slime mold {\it Physarum polycephalum} to adapt its locomotion to periodic changes of the environment~\cite{saigusa08a}.
The simplicity of the system - a unicellular organism - and its well-defined response to
specific input signals, make it an ideal test bed for the application of the notion of memory circuit elements in biology, and a source of inspiration for more complex adaptive behavior in living organisms. In addition, this is
a particularly appealing example of the full range of properties of memelements, in particular, their {\it analog} capability, which expands their range far beyond the digital domain.

In particular, it has been shown in a recent experiment~\cite{saigusa08a} that the {\it Physarum polycephalum} subjected to periodic unfavorable environmental conditions (lower temperature and humidity) not only responds to these conditions by reducing its locomotion speed, but also anticipates future unfavorable environmental conditions reducing the speed at the time when the next unfavorable episode would have occurred. While the microscopic mechanism of such behavior has not been identified by the authors of that work~\cite{saigusa08a}, their experimental measurements clearly prove the ability of {\it Physarum polycephalum} to anticipate an impending environmental change.

More specifically, the locomotion speed of the {\it Physarum polycephalum} was measured when favorable environmental  conditions (26$^\circ$C and 90$\%$ humidity) were interrupted by three equally spaced 10 min pulses of unfavorable environmental conditions (22$^\circ$C and 60$\%$ humidity)~\cite{saigusa08a}. The time separation between the pulses $\tau$ was selected between 30 and 90 minutes. It was observed that the locomotion speed at favorable conditions (approximately 0.25 mm/10min as shown in Fig. 1  of Ref. \cite{saigusa08a}) turns to close to zero each time the unfavorable conditions were presented. However, {\it spontaneous in-phase slow downs} were observed after time intervals $\tau$, $2\tau$ and even $3\tau$ after the last application of unfavorable conditions. In addition, if -- after a long period of favorable conditions -- a single pulse of unfavorable conditions is applied again then a spontaneous slow down (called a {\it spontaneous in-phase slow down after one disappearance}~\cite{saigusa08a}) after a time interval $\tau$ was observed. It clearly follows form this experiment that the {\it Physarum polycephalum} has a mechanism to memorize ("learn") the periodicity of environmental changes and adapts its behavior in anticipation of next changes to come~\cite{saigusa08a}.

\begin{figure}[bt]
\centering \includegraphics[width=8cm]{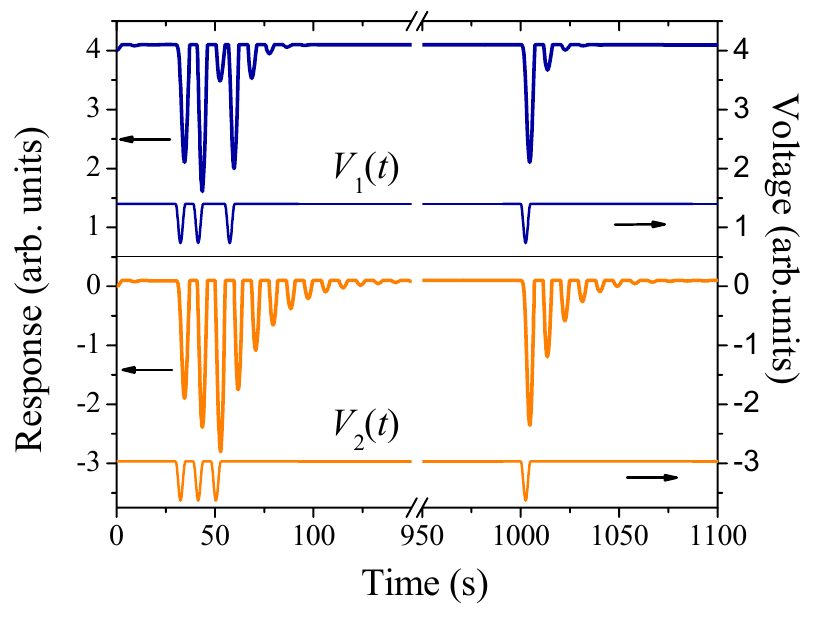}
\caption{ Modeling of the spontaneous in-phase slow-down responses~\cite{pershin09b}. This plot demonstrates that stronger and
longer-lasting responses for both spontaneous in-phase slow down and spontaneous in-phase slow down after one disappearance of the
stimulus are observed only when the circuit was previously trained by a periodic sequence of three equally spaced pulses as present in
$V_2(t)$. The applied voltage $V_1(t)$ is irregular and thus the three first pulses do not "train" the circuit.
Reprinted with permission from Ref. \cite{pershin09b}. \copyright  2009 American Physical Society.}
\label{fig:3}       
\end{figure}

We have developed a circuit model~\cite{pershin09b} of the adaptive behavior of the slime mold which has been later realized experimentally~\cite{driscoll10a} using vanadium dioxide as memory element~\cite{driscoll09b,driscoll09a}. The learning circuit is shown in Fig. \ref{fig:2}(a). Here, the role of environmental conditions is played by the input voltage $V(t)$ and the speed of locomotion is mapped into the voltage across the capacitor $C$. The learning circuit design resembles a damped $LC$ contour in which the amount of damping is controlled by the state of the memristive system $M$. To understand the circuit operation, we note that the memristive system employed in the circuit is of a threshold type (see Eqs. (\ref{eq3}) and~(\ref{eq4})), namely, its state can be significantly changed only by a voltage (across the memristive system) with a magnitude exceeding a certain threshold. Fig. \ref{fig:2}(b) presents the switching function $f(V)$ used to describe a threshold-type memristive system.

Our simulations of the learning circuit response to irregular and regular sequences of pulses are shown in Fig. \ref{fig:3}.
In these simulations, the scheme described above has been used with the only restriction that
the response signal cannot exceed a certain value~\cite{pershin09b} (electronically, a cut-off can be easily obtained by using an additional resistor and diode). When an irregular sequence of pulses is applied to the circuit, the voltage oscillations across the capacitor can not exceed the threshold voltage of the memristive system $M$ which continues to stay in its initial low-resistance state, thus damping the circuit. When the pulses are applied periodically with a period close to the period of the $LC$ contour oscillations, a sufficiently strong voltage across the capacitor $C$ is induced. This voltage switches the memristive system into the high-resistance state. Therefore, in this case, oscillations in the contour are less damped and last longer as Fig. \ref{fig:3} demonstrates. These oscillations exactly model the spontaneous in-phase slow down and in-phase slow down after one disappearance effects observed experimentally~\cite{saigusa08a}. We note that a single learning circuit memorizes past events of a frequency close to the resonance frequency of $LC$ contour. An array of learning circuits would model the learning of {\it Physarum polycephalum} in the whole frequency range~\cite{pershin09b}.

\begin{figure}[tb]
\centering \includegraphics[width=10cm]{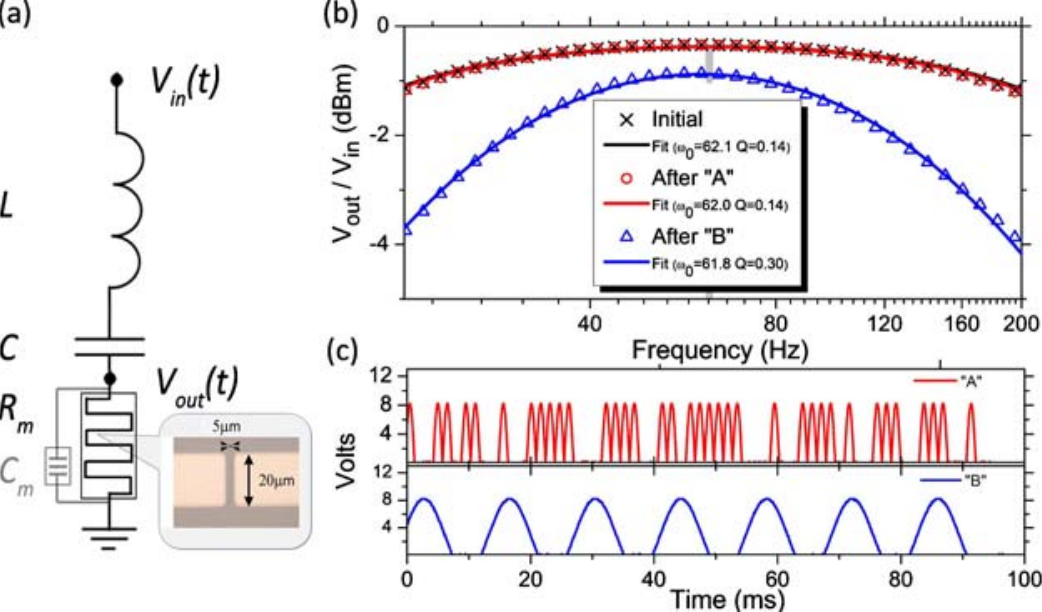}
\caption{Experimental realization of the learning circuit (adaptive filter) based on vanadium dioxide memristive device. (a) Schematic of the adaptive filter in which the memristive device (with a small memcapacitive component) is connected in series with a capacitor $C$ and inductor $L$. We note that such realization of the learning circuit operates similarly to the learning circuit shown in Fig.~\ref{fig:2}(a). (b) Small-signal (10 mV) transfer function ($V_{out} /V_{in}$) for the adaptive filter plotted before
and after off-resonance "A" and on-resonance "B" pulses. Solid lines are $RLC$ bandpass-filter fit to data,
which generates the $\omega_0$ and $Q$ values in the legend. Pulse sequence B has a significant training effect on the
circuit, while A has little or no effect. (c) Time series of the off-resonance "A" sequence of pulses and on-resonance
"B" sequence of pulses.  Reprinted with permission from Ref. \cite{driscoll10a}. \copyright  2010 American Institute of Physics.}
\label{fig:4}       
\end{figure}

Recently, an experimental implementation of this learning circuit has been reported~\cite{driscoll10a}. In this work, a learning circuit
similar to that in Fig. \ref{fig:2}(a) has been built with the only difference that the memristive system (a vanadium dioxide memristive device~\cite{driscoll10a}) has been connected in series with a capacitor (Fig. \ref{fig:4}(a)). The memristive properties of vanadium dioxide are based on an insulator-to-metal phase transition occurring in this material in the vicinity of 342K~\cite{driscoll09b,driscoll09a}. In order to realize the memristive functionality, the vanadium dioxide device is heated to a steady-state temperature of 339.80 K (right below the transition temperature) and subjected to an externally applied voltage. The Joule heating (due to the applied voltage) incrementally drives the vanadium dioxide material through the phase transition, thus reducing its resistance. The operation of the learning circuit depicted in Fig. \ref{fig:4}(a) is then clear. While off-resonance signals applied to the circuit can not excite a sufficient current to drive the vanadium dioxide through the phase transition, the current generated by resonance signals is sufficient for this purpose. Fig. \ref{fig:4}(b) demonstrates modification of the transfer function of the circuit caused by off-resonance and resonance pulse sequences (Fig. \ref{fig:4}(c)) applied to the circuit. Fig. \ref{fig:4}(b) clearly indicates a change in the transfer function caused by resonance signals (learning).

Finally, we mention that the formalism of memory circuit elements~\cite{diventra09a} has also been useful in modeling biophysical systems whose electric response depends on the history of applied voltages or currents. An example of such situation is the electro-osmosis in skin which has been recently described by a memristive model~\cite{Johnsen11a}. Physically, the voltage applied to the skin induces a water flow in sweat-duct capillaries changing the skin conductance. The position of the water table (the level separating dry and wet zones) in capillaries plays the role of the internal state variable whose dynamics is determined by the applied voltage~\cite{Johnsen11a}. The memristive model of electro-osmosis~\cite{Johnsen11a} is in a good agreement with experimental data and further demonstrates the potential of the formalism of memory circuit elements for modeling biophysical processes.

\subsection{Neuromorphic circuits} \label{sec:32}

Our second example of biologically-inspired circuits with memory circuit elements is from the area of neural networks.
Neural networks form a class of circuits whose operation mimics the operation of the human (and animal) brain.
Below, we consider electronic implementations of two important processes occurring in biological neural networks:
associative memory and spike timing-dependent plasticity. Both features can be implemented in artificial neural
networks based on memristive synapses.

\subsubsection{Associative memory}

The associative memory is one of the most fundamental functionalities of the human (and animal) brain.
By making associations we learn, adapt to a changing environment and better retain and recall events. One of the most famous
experiments related to associative memory is Pavlov's experiment~\cite{pavlov27a} whereby salivation of a dog's
mouth is first set by the sight of food. Then, if the sight of food is accompanied by a sound (e.g., the tone of a bell) over a
certain period of time, the dog learns to associate the sound to the food, and salivation can be triggered by the sound alone,
without the intervention of vision.

Recently, we have reproduced~\cite{pershin10c} the Pavlov's experiment utilizing a neural network with memristive synapses.
As a first example, we have implemented the well known Hebbian rule introduced by Hebb in 1949:
"when an axon of cell A is near enough to excite a cell B and repeatedly or persistently takes part in firing it,
some growth process or metabolic change takes place in one or both cells such that A's efficiency, as one of the cells firing B, is increased"~\cite{Hebb49a}.
To put it differently, the neurons that fire together, wire together. 

\begin{figure}[tb]
\sidecaption
\includegraphics[width=7cm]{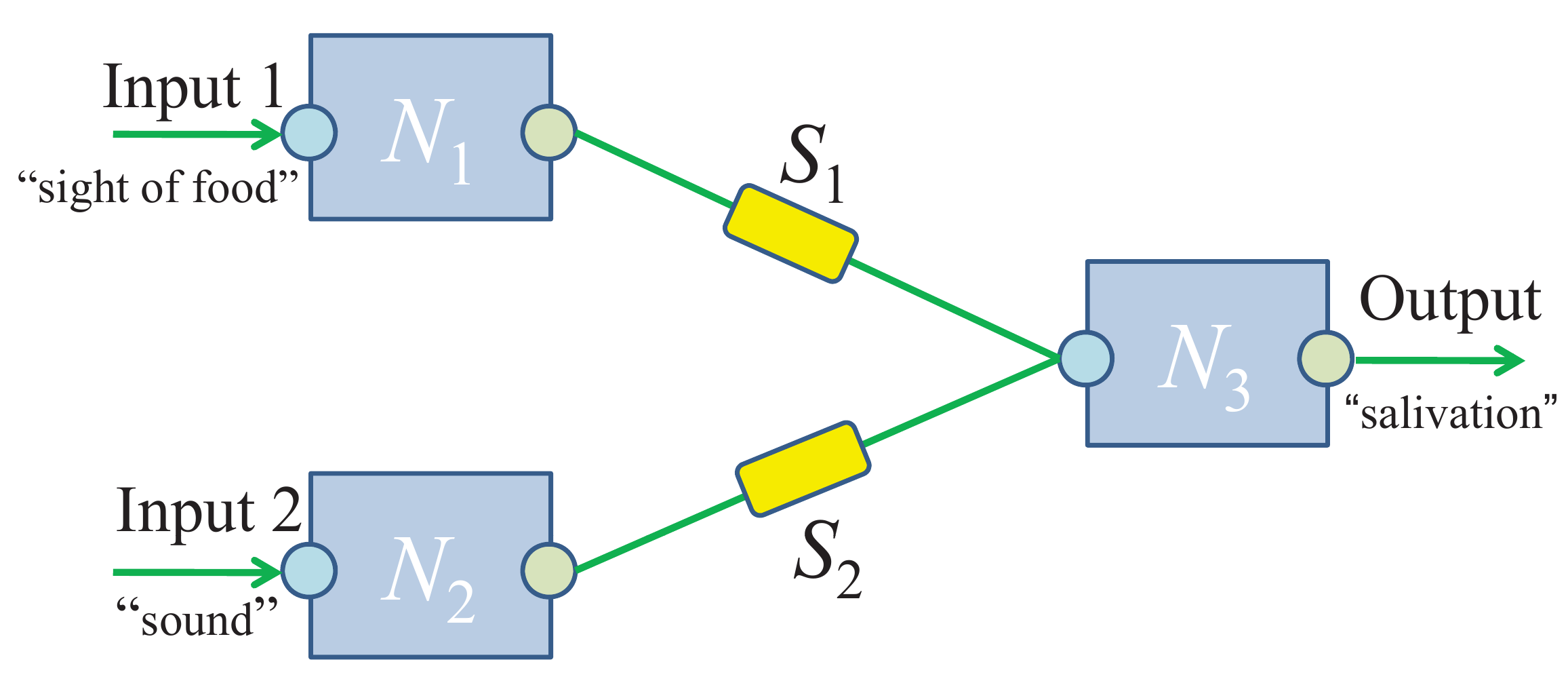}
\caption{Memristive neural network with an associative memory ability. Here, two input neurons ($N_1$ and $N_2$) are connected through memristive synapses ($S_1$ and $S_2$) to the output neuron $N_3$. The details of circuit operation are given in the text. Reprinted from Ref. \cite{pershin10c}. \copyright  2010 with permission from Elsevier.}
\label{fig:5}       
\end{figure}

In order to show associative memory, let us consider a simple neural network consisting of three electronic neurons and two memristive synapses as shown in Fig. \ref{fig:5}. We assume that the first input neuron activates under a specific ("visual") event,
such as "sight of food", and the second input neuron
activates under another ("auditory") event, such as a particular "sound".

\begin{figure}[b]
\centering \includegraphics[width=7cm]{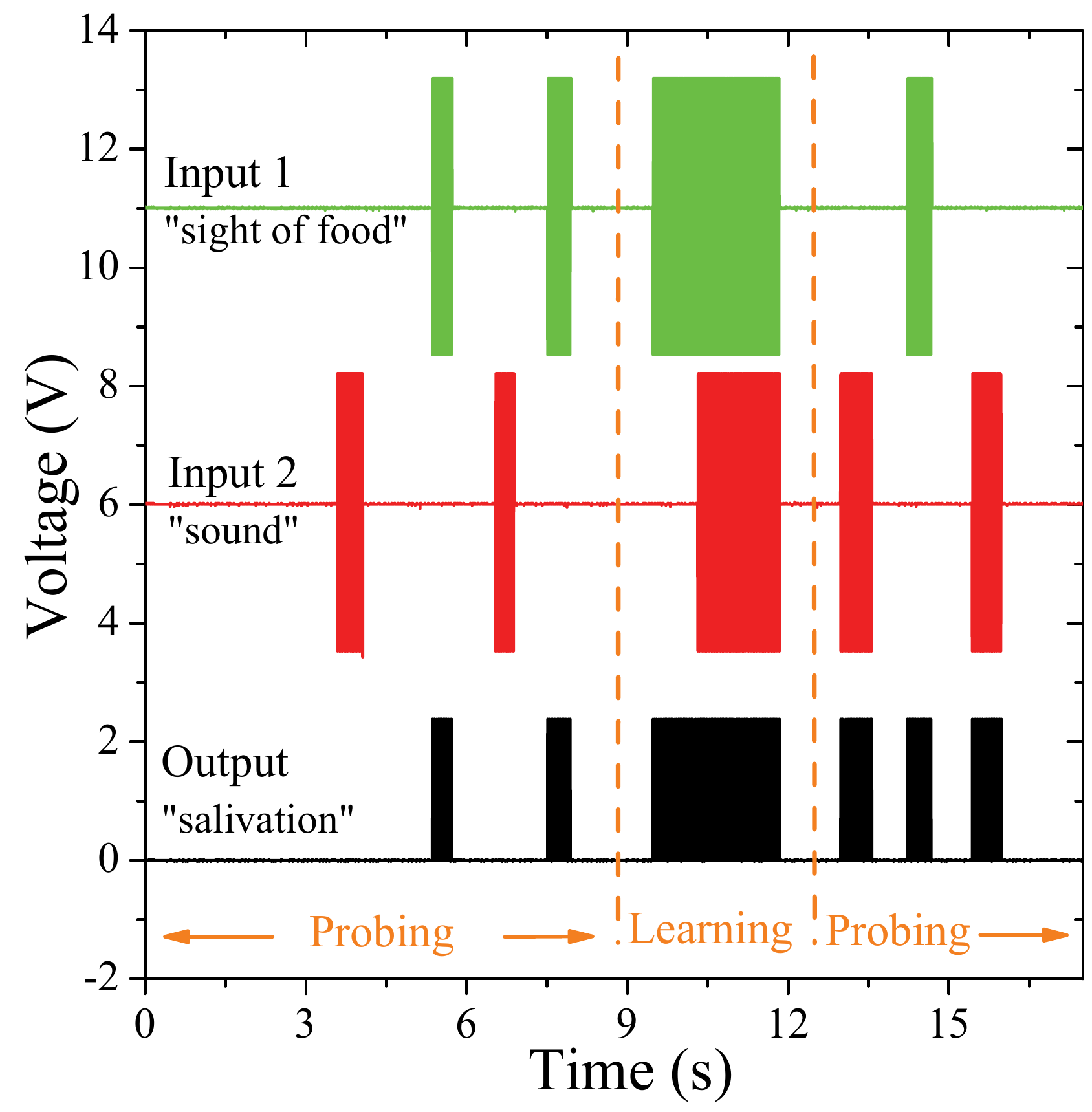}
\caption{Experimental demonstration of the associative memory with
memristive neural networks. In this experiment, a simple neural network shown in Fig. \ref{fig:5}
was realized. The first "probing" phase demonstrates that, initially, only a signal from N$_1$ neuron activates
the output neuron. The association of the Input 2 signal with the Output develops in the "learning phase" when
N$_1$ and N$_2$ neurons are simultaneously activated. In this case, a signal at the Input 1
excites the third neuron that sends back-propagating pulses of a negative
amplitude. These pulses, when applied simultaneously with forward propagating pulses from the Input 2 to the second memristive synapse S$_2$ cause it to learn. The final "probing" phase demonstrates that signals from both N$_1$ and N$_2$ activate the output neuron.
A detalied description of the experiment is given in Ref.  \cite{pershin10c}. Reprinted from Ref. \cite{pershin10c}. \copyright  2010 with permission from Elsevier.}
\label{fig:6}       
\end{figure}

On the electronic level, an electronic neuron sends forward (to its output) and backward (to its input) voltage spikes of opposite polarity when the amplitude of the input signal exceeds a threshold value. Regarding the dynamics of memristive synapses, they have been selected of a threshold-type (Eqs. (\ref{eq3}) and~(\ref{eq4})) with a threshold voltage exceeding the output voltage of electronic neurons. In this case, voltage spikes applied to a single terminal of a memristive synapse is not enough to induce its change. The latter is possible only if forward and backward propagating spikes overlap in time across a synapse. We have employed memristor emulators~\cite{pershin10d,pershin10c} as memristive synapses\footnote{Several designs of memristor~\cite{pershin10d,biolek11b,biolek11a,chua71a} as well as memcapacitor and meminductor~\cite{biolek11b,biolek10a,pershin11b,pershin09e} emulators are known in the literature. These emulators serve as an important practical tool to build small-scale circuits with memory circuit elements.}. The main components of a memristor emulator are a digital potentiometer, a  microcontroller and an analog-to-digital converter. Using the converter, the microcontroller continuously reads the voltage applied to the digital potentiometer and updates the potentiometer resistance according to a pre-programmed model of a voltage- or current-controlled memristive system. The operation of electronic neurons is realized along similar lines~\cite{pershin10c}.  Operation of the associative memory is presented in Fig.~\ref{fig:6} where a detail of this process is given.

Our work as described above~\cite{pershin10c} demonstrates the potential of memristive devices for neuromorphic circuits applications. Importantly, it has been recently shown in numerous experiments that memristive devices can be built at the nanoscale~\cite{tamura06a,waser07a,lee07a,Dietrich07a, Raoux08a,inoue08a,strukov08a,Sawa08a,jo09a,Schindler09b,jo10a,pershin11a,Lee11a,Torrezan11a}.
This opens up the possibility to fabricate neuromorphic circuits with the same amount of memristive synapses as the number of biological synapses in the human brain ($\sim 10^{14}$). In fact, one of the main challenges for practical realizations of an artificial brain on a chip is related to the high connectivity of biological neurons. It has been estimated that, on average, the number of connections per neuron is of the order of $10^3$. Therefore, neural networks of memelements built at the nanoscale offer advantages - in terms of density - unavailable with current active elements (such as transistors).

\subsubsection{Spike timing-dependent plasticity}

However, the above mentioned simple Hebbian rule does not describe the much more complicated time-dependent plasticity of biological synapses~\cite{Levy83a,Markram97a,Bi98a,Froemke02a}. The latter has come to be known as spike timing-dependent plasticity (STDP). When a post-synaptic signal reaches the synapse {\it before} the action potential of the pre-synaptic neuron, the synapse shows long-term depression (LTD), namely its strength decreases (smaller connection between the neurons) depending on the time difference between the post-synaptic and the pre-synaptic signals. Conversely, when the post-synaptic action potential reaches the synapse {\it after} the pre-synaptic action potential, the synapse undergoes a long-time potentiation (LTP), namely the signal transmission between the two neurons increases
in proportion to the time difference between the pre-synaptic and the post-synaptic signals. The learning process and the storing of information in the brain thus follow non-trivial time-dependent features which have not
been fully understood yet. Implementation of STDP in artificial networks can thus help unraveling these mechanisms.

\begin{figure}[tb]
\sidecaption
\includegraphics[width=6cm]{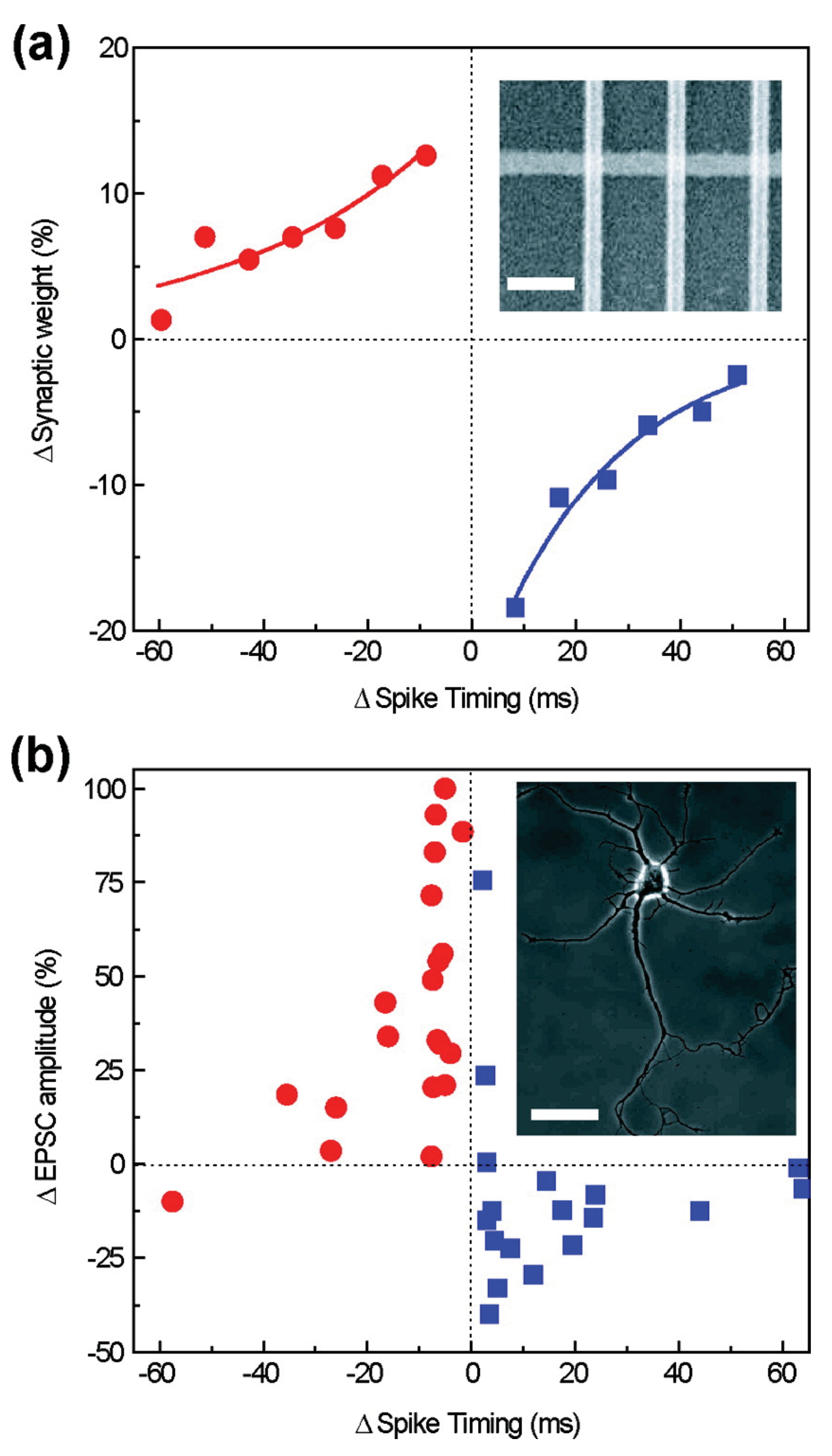}
\caption{(a) Measured change in the synaptic weight versus spike separation. Inset: SEM image of
the memristive crossbar array, scale bar is 300 nm. (b) Measured change in excitatory postsynaptic current of rat
neurons after repetitive correlated spiking versus relative spiking timing (the plot was reconstructed
from Ref. \cite{Bi98a}). Inset: image of a hippocampal neuron (the image was adapted
with permission from reference \cite{Kaech06a}). Scale bar is 50 $\mu$m. Reprinted with permission from \cite{jo10a}. Copyright 2010 American Chemical Society.}
\label{fig:STDP}       
\end{figure}

 The spike timing-dependent plasticity can be implemented using different types of memristive systems. Following our previous work~\cite{pershin10e}, neuromorphic circuits can be based on memristive systems with or without an internal spike-timing tracking capability. In the most simple case, memristive systems without spike-timing tracking capability are of the first order, while those supporting such capability are of the second order as an additional internal state variable is needed to track timing of pre-synaptic and post-synaptic pulses~\cite{pershin10e}.
 In the first case, an additional external hardware is required to implement the spike timing-dependent plasticity. For example, STDP was recently realized using a combination of memristive systems with CMOS (complementary metal-oxide-semiconductor) elements \cite{jo10a} (see Fig. \ref{fig:STDP}). Another approach involves utilization of overlapping pulses of opposite polarities~\cite{pershin10e,snider08a,parkin10a}.

A simple second-order memristive system with time-tracking capability is described by the following equations~\cite{pershin10e}
\begin{eqnarray}
R&=&x, \label{stdp_memr0} \\ \dot x&=&\gamma \left[ \theta (V_M-V_t) \theta (y-y_t) +  \theta (-V_M-V_t) \theta (-y-y_t) \right]y,    \label{stdp_memr1} \\
\dot y&=&\frac{1}{\tau} \left[ -V_M \theta (V_M-V_t) \theta (y_t-y) -  V_M\theta (-V_M-V_t) \theta (y+y_t) -y \right],
 \label{stdp_memr2}
\end{eqnarray}
where $x$ and $y$ are internal state variables, $\gamma$ is a constant, $V_t$ is a threshold voltage, $y_t$ is the threshold value of $y$, and $\tau$ is a constant defining the time window of STDP. The second-order memristive system with timing tracking capability defined by the above equations is very promising for neuromorphic circuits application since neuron's firing can be implemented simply  by short single rectangular pulses and no additional hardware as in the case of first-order memristors (see, e.g., Ref. \cite{jo10a}). However, such solid-state second-order memristors need to be developed, even though their implementation in memristor emulators~\cite{pershin10d,biolek11b} can be easily realized.

\begin{figure}[tb]
 \begin{center}
  \centerline{
      \mbox{(a)}
    \mbox{\includegraphics[width=5.50cm]{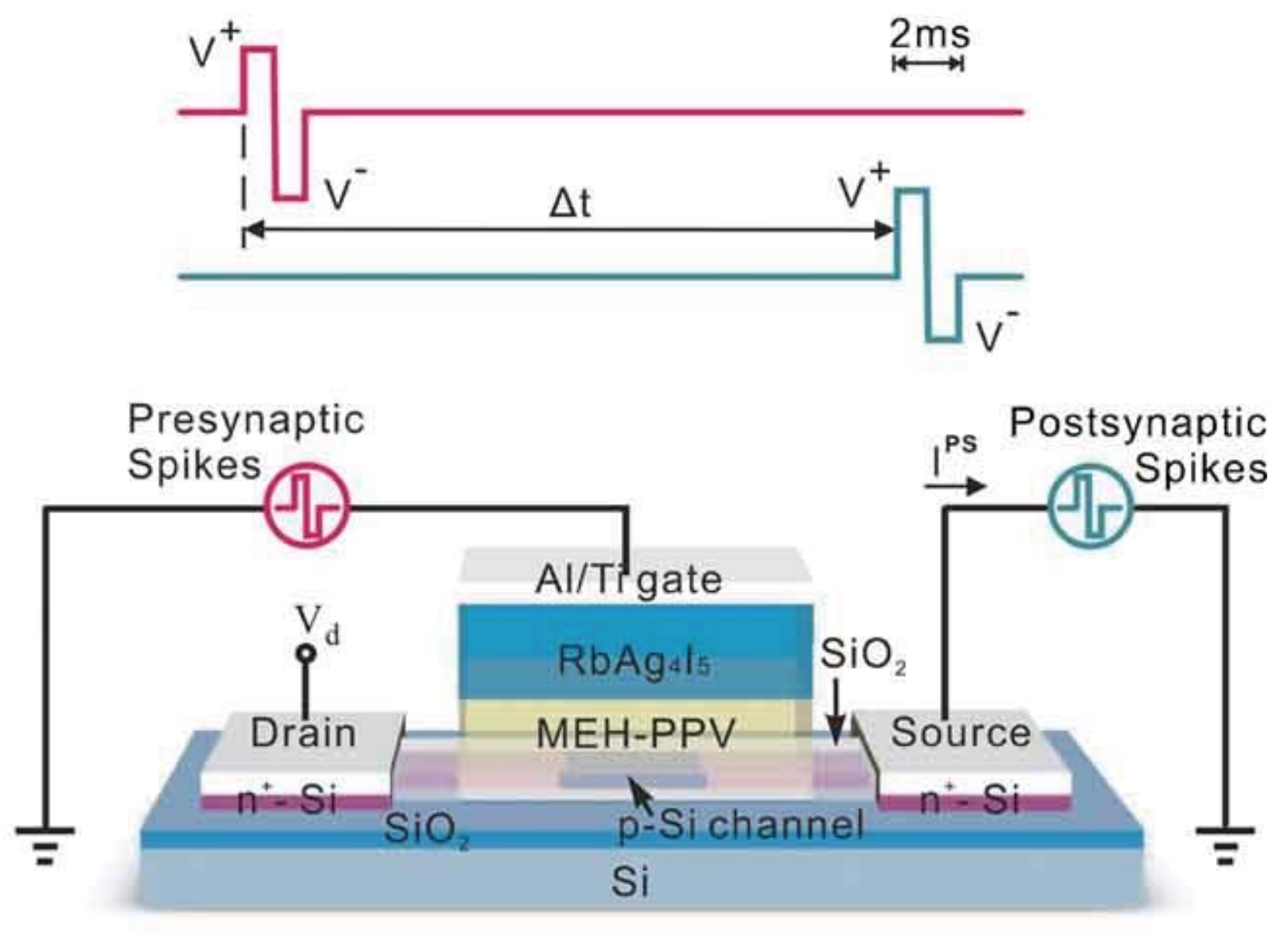}}
        \mbox{(b)}
    \mbox{\includegraphics[width=5.50cm]{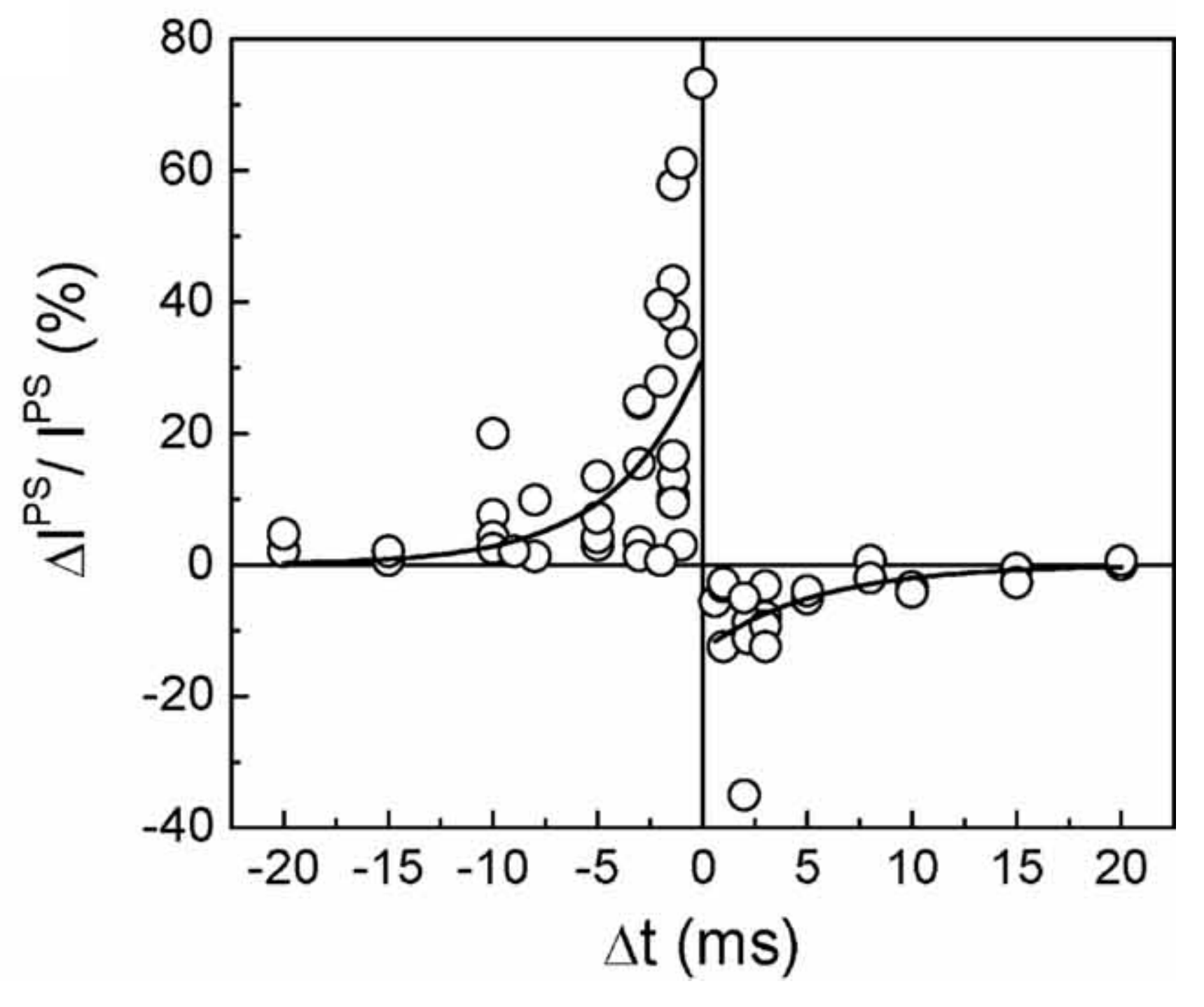}}
   }
\caption{ (a) Structure of synaptic transistor . (b) The relative
changes of the postsynaptic currents measured after application of
120 pairs of temporally correlated pre- and post-synaptic spikes. From \cite{Lai10a}, copyright Wiley-VCH Verlag GmbH $\&$ Co. KGaA.
Reproduced with permission.} \label{fig:Lai}
\end{center}
\end{figure}

Moreover, several authors have discussed applications of three-terminal transistor-like electronic devices with memory~\cite{Alibart10a,Lai10a,Zhao10a} in the area of neuromorphic computing. Although, formally  such devices can not be categorized as memristive systems, their operation is clearly based on memristive effects. In particular, Lai {\it et al.}~\cite{Lai10a} have experimentally fabricated a synaptic transistor. For instance, Fig. \ref{fig:Lai} depicts their experimental scheme and selected measurement results that confirm realization of spike timing-dependent plasticity in this device.

\subsection{Networks of memory circuit elements} \label{sec:33}

A human brain - and also the brain of many other living organisms - solves many problems much better than traditional computers. The reason
for this is a type of massive parallelism in which a large number of neurons and synapses participate simultaneously in the
calculation. Here, we consider networks of memory circuit elements and their ability to {\it i)} solve efficiently certain graph theory optimization problems, and {\it ii)} retain such information for later use. In particular, we demonstrate that a network of memristive devices solves the maze problem much faster than any existing algorithm~\cite{pershin11d}. Similar to the brain's operation, such an extraordinary advance in computational power is due to the massively-parallel network dynamics in which all network components are simultaneously involved in the calculation. This type of parallelism could be dubbed as an {\it analog parallelism} which is
very different from that used in conventional supercomputers. In the latter systems, each core typically runs a separate process that, relatively rarely, exchanges information with other cores. In calculations done by networks, the information exchange is continuous resulting in a tremendous increase of computational power as we demonstrate below.

\begin{figure}[tb]
\centering \includegraphics[width=12cm]{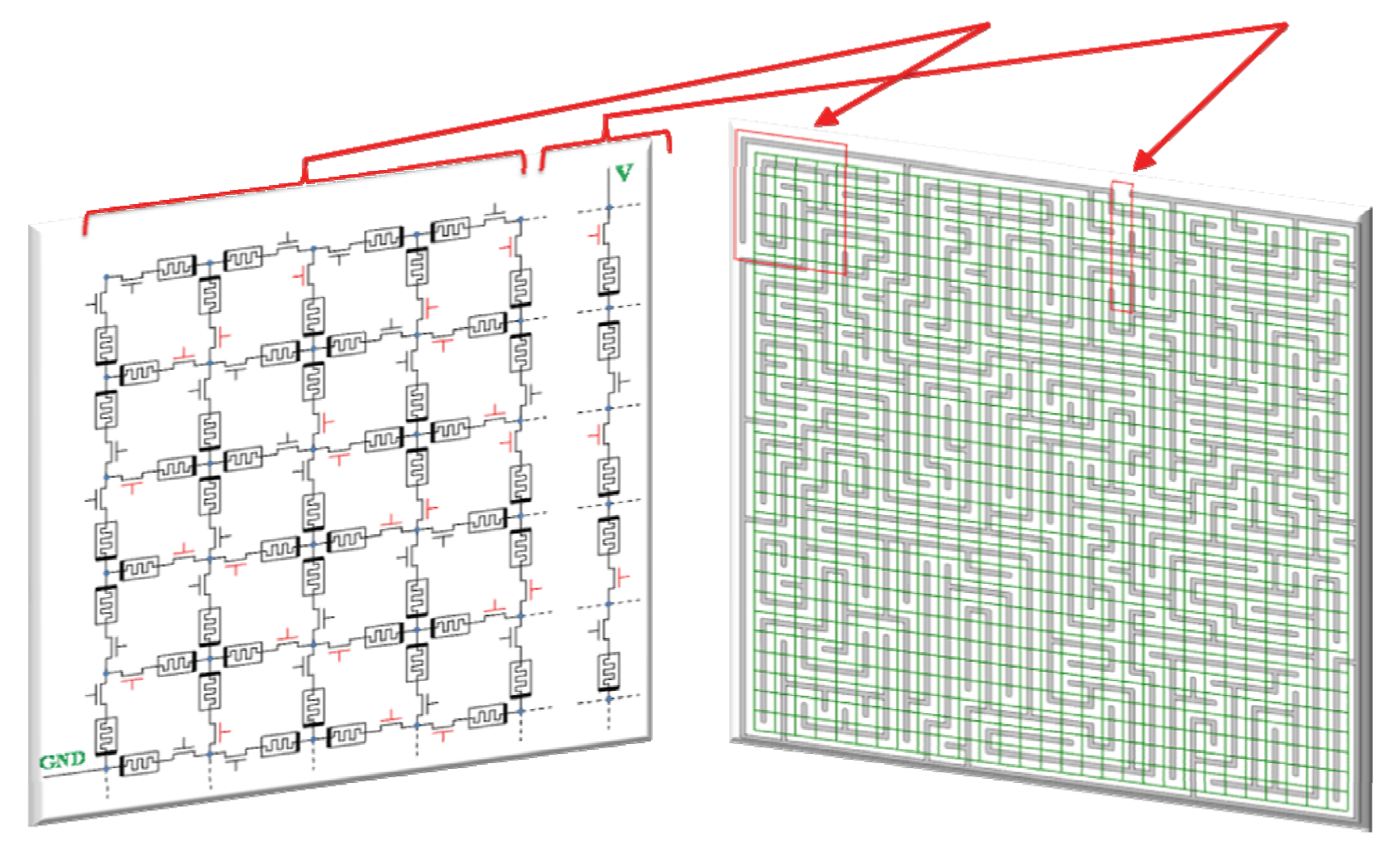}
\caption{Maze mapping into a network of memristors. {\it Right panel}. The maze is covered by an array of vertical and horizontal lines having the periodicity of the maze. {\it Left panel}. Architecture of the network of memristors in which each crossing between vertical and horizontal lines in the array (in the right panel) is represented by a grid point to which several basic units consisting of memristors and switches (field-effect transistors) are linked in series. The maze topology is encoded into the state of the switches such that if the short line segment connecting neighboring crossing points in the array crosses the maze wall then the state of the corresponding switch is "not connected" (shown with red symbols). All other switches are in the "connected" state. The external voltage ($V$) is applied across the connection points corresponding to the entrance ($V$) and exit (ground, $GND$) points of the maze.  Reprinted with permission from Ref. \cite{pershin11d}. \copyright  2011 American Physical Society.}
\label{fig:7}       
\end{figure}

Left panel of Fig. \ref{fig:7} depicts a memristive network (memristive processor) in which points of a square grid are connected
 by basic units (memristive system plus switch (FET))~\cite{pershin11d}.  Each switch in the network can be in the "connected" or "not-connected" state. Since the direction of current flow in the network is not known {\it a priori}, the polarity of adjacent memristive devices (indicated by the black thick line in the memristor symbol in Figure \ref{fig:1}) is chosen to be alternating. Experimentally, the suggested network could be fabricated using, e.g., CMOL (Cmos+MOLecular-scale devices) architecture \cite{Likharev05a} combining a single memristor layer with a conventional CMOS layer.
The operation of the massively-parallel processor consists of three main stages: initialization, computation and reading out the computation result. All these stages are realized by externally applied signals (originating, e.g., from the CMOS layer).

\begin{figure}[tb]
\centering \includegraphics[width=8cm]{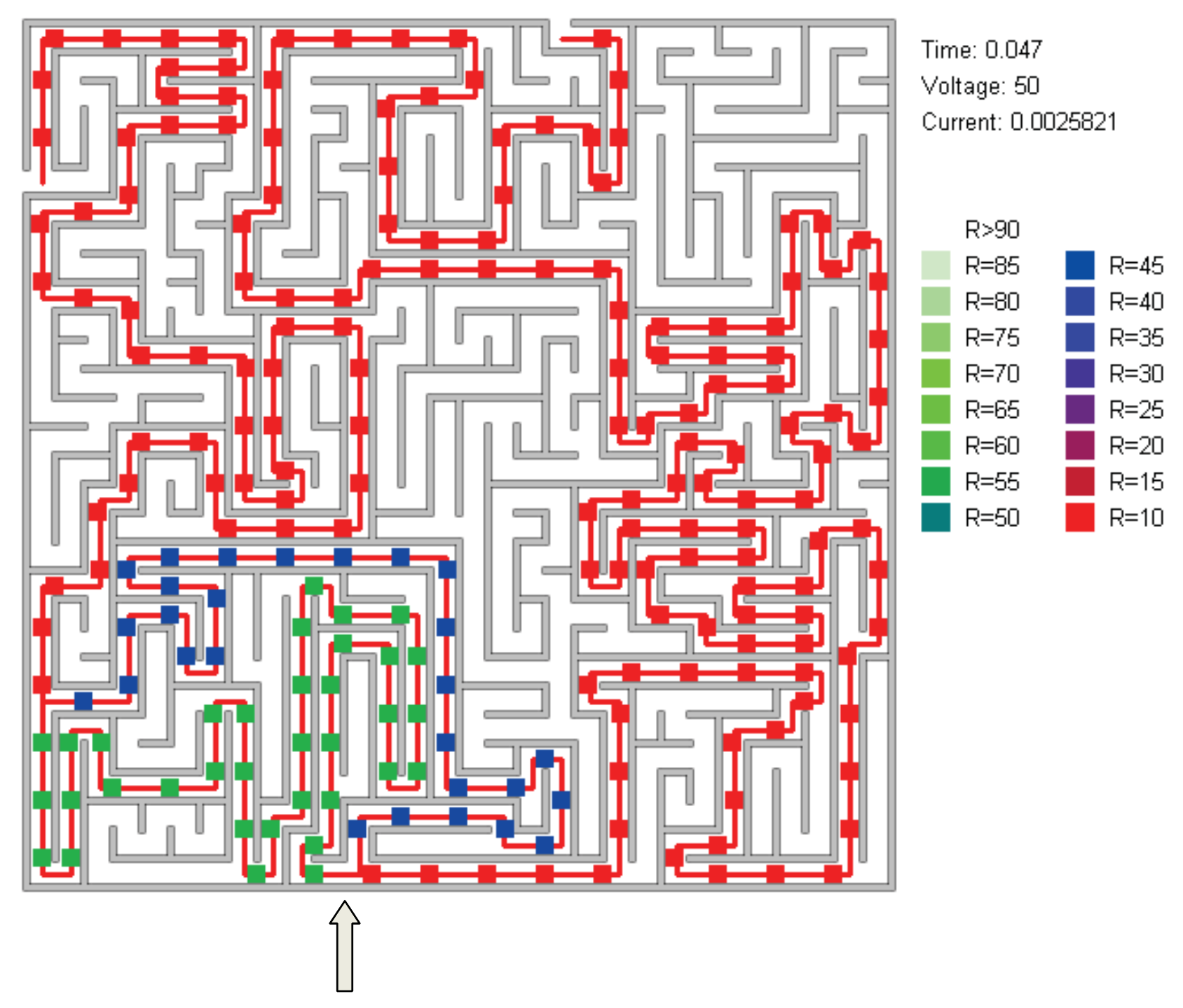}
\caption{Solution of a multiple-path maze~\cite{pershin11d}.
The maze solution contains two common segments (red  dots connected by a red line), and two alternative segments
of different lengths close to the left bottom corner. The memristance in the shorter segment (blue  dots connected by a red  line) is smaller than that in the longer segment (green dots connected by a red  line)
since the current through the shorter segment is larger and,
consequently, the change of the memristors' state along this
segment is larger. The arrow at the bottom indicates a splitting point of the solution path. The resistance is in
 Ohms, the voltage is in Volts and the current is in Amperes. Reprinted with permission from Ref. \cite{pershin11d}. \copyright  2011 American Physical Society.}
\label{fig:8}       
\end{figure}

During the first initialization stage, all memristive elements in the network are switched into the "OFF" state. This can be done, for example, by applying $GND$ and appropriately selected $V_1$ voltages in a chessboard-like pattern to all grid points of the memristive network for a sufficiently long period of time~\cite{pershin11d}. After that, the maze topology is mapped onto the memristive network by setting appropriate switches into the "not connected" state. We describe this process in the caption of Fig. \ref{fig:7}. The computation stage consists in the application of a single voltage pulse of appropriate amplitude and duration across grid points corresponding to the entrance and exit points of the maze. The solution can be later read or used in further calculations.

We have modeled the memristive processor operation by numerically solving Kirchhoff's current law equations complemented by Eqs. (\ref{eq1struk}), (\ref{eq2struk}) which in the present network case are modified as
\begin{equation}
R_{ij}^M=R_{ON}x_{ij}+R_{OFF}\left( 1 -x_{ij}\right),\label{eq1ij}
\end{equation}
where $R_{ON}$ and $R_{OFF}$ are again the minimal and maximal values of memristance, $x_{ij}$ is the dimensionless internal state variable for each memristor bound to the region  $0\leq x_{ij} \leq 1$, and ($i,j$) are grid indexes of a memristor to identify its location in the network. The dynamics of $x_{ij}$ is then given by
\begin{equation}
\frac{\textnormal{d} x_{ij}}{\textnormal{d} t}=\alpha I_{ij}(t) \label{eq2ij},
\end{equation}
with $\alpha$ a constant and $I_{ij}(t)$  the current flowing through the memristor ($ij$).

Fig. \ref{fig:8}(a) shows a solution of a multiple-path maze. The maze solution is clearly seen in Fig. \ref{fig:8} as chains of red, blue and green boxes (representing memristive devices with lower memristance) connected by a red line. Importantly, the memristive processor not only determines all possible solutions of the maze but also stores them and sorts them out according to their length. This feature is described in more details in the caption of Fig. \ref{fig:8}. Also, the memristive processor requires only {\it one single step} to find the maze solution thus outperforming all known maze solving approaches and algorithms.

We also note that the wide selection of physical mechanisms of memory we can "shop" from, offers many opportunities to design novel efficient electronic devices \cite{pershin11a}. For example, a memristive processor based on fast switching nanoionic metal/insulator/metal cells~\cite{pershin11a} would require just few nanoseconds or even less~\footnote{Fast sub-nanosecond switching has been recently reported in tantalum oxide memristive systems~\cite{Torrezan11a}.} to solve the maze. More generally, a network of memristors - or other memory circuit elements - can be considered as an adaptable medium whose state dynamically changes in response to time-dependent signals or changes in the network configuration. Therefore, the use of these processors is
not limited to maze solving: We expect they could help find the solution of many graph theory optimization problems including, for instance, the traveling salesman problem, shortest path problem, etc.

\section{Conclusions and Outlook} \label{sec:4}

In conclusion, we have shown that the two-terminal electronic devices with memory -- memristive, memcapacitive and meminductive systems -- are very useful to model a variety of biological processes and systems. The electronic
implementation of all these mechanisms can clearly lead to a novel generation of "smart" electronic circuits that can find useful applications in diverse areas of science and technology. In addition, these
memelements and their networks, provide solid ground to test various hypothesis and ideas regarding the functioning of the human (and animal) brain both theoretically and experimentally. Theoretically because their
flexibility in terms of what type and how many internal state variables responsible for memory, or what network topology are required to reproduce certain biological functions can lead to a better understanding of the microscopic
mechanisms that are responsible for such features in living organisms. Experimentally because with the continuing miniaturization of electronic devices, memelements can be assembled into networks with similar densities as
the biological systems (e.g., the brain) they are designed to emulate. In particular, we anticipate potential applications for memcapacitive and meminductive systems~\cite{diventra09a} which offer such an important property as low energy dissipation combined with information storage capabilities. We are thus confident that the area of biologically-inspired electronics with memory circuit elements will offer many research opportunities in several fields of science and technology.

\begin{acknowledgement}
M.D. acknowledges partial support from the NSF Grant No. DMR-0802830.
\end{acknowledgement}

\bibliographystyle{spphys}
\bibliography{maze}
\end{document}